\documentclass{PoS}

\title{4.5-year simultaneous multi-wavelength observation of Mrk 421 in the ARGO-YBJ and Fermi overlap era}

\ShortTitle{multi-wavelength observation of Mrk 421}

\author{Songzhan Chen\thanks{This work is supported in China by NSFC  No.11205165}\\
        Key Laboratory of Particle Astrophysics, Institute
                  of High Energy Physics, CAS,
                  Beijing 100049,  China.\\
        E-mail: \email{chensz@ihep.ac.cn}}

\author{\speaker{Silvia Vernetto} \\
        Osservatorio Astrofisico di Torino - INAF - Italy\\
 Istituto Nazionale di Fisica Nucleare - Sezione di Torino - Italy}
\author{On behalf of the ARGO-YBJ Collaboration}

\abstract{As one of the most active blazars, Mrk421 is an excellent candidate for the study of the physical processes within the jets of AGN. Here we report on the extensive multi-wavelength observations of Mrk 421 over 4.5 years, from 2008 August to 2013 February. This source was simultaneously monitored by several experiments at different wavelengths:  ARGO-YBJ in TeV $\gamma$-rays, $Fermi$-LAT in GeV $\gamma$-rays, $Swift$-BAT in hard X-rays, $RXTE$-ASM, $MAXI$ and $Swift$-XRT in soft X-rays, $Swift$-UVOT in ultraviolet, and OVRO in radio frequencies.
In particular, thanks to the ARGO-YBJ and $Fermi$ data, the whole energy range from 100 MeV to 10 TeV is covered without any gap.
 According to the observed light curves, ten states (including seven large flares, two quiescent phases and one outburst) were selected. For the first time, the multi-wavelength spectral evolutions of Mrk 421 during different states were systematically analyzed.
During the outburst phase and the seven flaring episodes, the peak energy in X-rays is observed to increase from sub-keV  to   few keV. The TeV $\gamma$-ray flux  increases up to 0.9$-$7.2 times the flux of the Crab Nebula.  The behavior of GeV $\gamma$-rays is found to vary depending on the flare, a feature that leads us to classify flares into three groups according to the GeV flux variation. Finally, the observed radiation spectra above 0.3 keV of different states can be reasonably described by a simple one-zone synchrotron self-Compton model. The underlying physical mechanisms responsible for different states may be related to the acceleration process or to  variations of the ambient medium. }

\FullConference{The 34th International Cosmic Ray Conference,\\
		30 July- 6 August, 2015\\
		The Hague, The Netherlands}

\begin{document}

\section{Introduction}
Mrk 421 (z=0.031), classified as a BL Lac object, is one of the brightest VHE $\gamma$-ray blazars known. It is a very active blazar with major outbursts  about once every two years, in both X-rays and $\gamma$-rays \cite{chen13}.
Actually it is considered  an excellent candidate to study  the physical processes within the AGN jets,
for which continuously multi-wavelength observation, particularly in
X-rays and VHE g-rays, are of
fundamental importance.
In the VHE band, Cherenkov telescopes cannot   regularly monitor AGNs, because of their limited duty
cycle and narrow field of view (FOV). The wide-FOV  EAS array ARGO-YBJ, with high duty cycles, is more suitable for this purpose. ARGO-YBJ
have continuously monitored the blazar Mrk 421 for 5 years, extending at higher energies the multi-wavelength survey carried out by the radio observatory OVRO, the satellite-borne X-ray detectors $Swift$, $RXTE$, $MAXI$ and the GeV $\gamma$-ray detector $Fermi$-LAT. In particular, thanks to the ARGO-YBJ and $Fermi$-LAT data, the high energy component of Mrk 421 SED has been completely covered without any gap from 100 MeV to 10 TeV. In this paper we report on the 4.5-year multi-wavelength data recorded from August 2008 to February 2013, a period that includes several large flares of Mrk 421. Such a long-term multi-wavelength observation is rare and provides a unique opportunity to investigate on the emission variability of Mrk 421 from radio frequencies to TeV $\gamma$-rays.

\section{Multi-wavelength observations and analysis}
The ARGO-YBJ experiment  is a full coverage extensive air shower detector made of RPCs, located at Yangbajing(4300 m a.s.l., Tibet, China).
The full detector has been in stable data taking from
2007 November to January 2013 with duty cycle   higher than $86\%$. Due to the full coverage configuration and the location at high altitude, the detector energy threshold is $\sim$300 GeV, much lower than any previous EAS array.
The data analysis is carried out as described in \cite{barto13a}.

$Fermi$-LAT is a pair-conversion telescope, with a FOV over 2 sr, active in the
100 MeV$-$300 GeV energy range with an unprecedented sensitivity.  $Fermi$-LAT  is monitoring the whole sky daily since August 2008.  In this work, we use the $Fermi$-LAT data from a region of radius $15^{\circ}$ centered on Mrk 421. The analysis is performed using the ScienceTools (Version v9r33p0) provided by
the $Fermi$-LAT collaboration  \footnote{http://fermi.gsfc.nasa.gov/ssc/}.
The data are processed following the standard recommendations for the binned method.

$Swift$-BAT   is a coded aperture mask imaging telescope (1.4 sr FOV)
operating since February 2005. The daily flux from Mrk 421 at energy 15$-$50 keV is provided by $Swift$-BAT\footnote{Transient monitor results provided by the $Swift$-BAT team:  http://heasarc.gsfc.nasa.gov/docs/swift/results/ \\ transients/weak/.}   and is used here to build the light curve.
To obtain the SED at 14$-$195 keV, we downloaded all the available data of Mrk 421 through the HEASARC data
archive\footnote{http://heasarc.nasa.gov/docs/archive.html}. The data analysis includes the recipes presented in \cite{ajello08,tueller10}.

The $MAXI$-GSC   detector, operating in the 2$-$20 keV range, started data taking in August 2009.
 The light curves at 2$-$20 keV for Mrk 421 are
publicly available \footnote{The MAXI data are provided by RIKEN, JAXA and the MAXI team: http://maxi.riken.jp/top/}.
 In this work we used them also to build the X-ray spectrum of Mrk 421, comparing the measured
counting rate in each band with the one of the Crab Nebula, used as a standard candle.

$RXTE$-ASM   consists of three proportional counters, each one with a field of
view of $6^{\circ}\times90^{\circ}$.
The $RXTE$-ASM data in the (2$-$12) keV range are publicly available \footnote{Quick-look results provided by the  $RXTE$-ASM team: http://xte.mit.edu/ASM\_lc.html.}, and were also used here to build the X-ray spectrum.
For Mrk 421, the daily flux is provided from 1995 up to the middle of 2010.

$Swift$-XRT  is a focusing X-ray telescope with energy range from 0.2 to 10 keV.
The XRT data in WT mode \footnote{http://heasarc.gsfc.nasa.gov/}  was processed with the XRTDAS software package (v.2.6.0) following the standard recommendations. The XRT average spectrum in the 0.3$-$10 keV energy band
was fitted using the XSPEC package (v.12.7.0). We adopted a LogParabolic model ($f(E)=J_{0} \cdot (E)^{-(a+b \cdot log(E)}$) for the photon-flux spectral density, with an absorption hydrogen-equivalent column
density fixed to the Galactic value in the direction of
the source, namely $1.92 \times 10^{20} $  $cm^{-2}$ for Mrk 421.

$Swift$-UVOT   is the ultraviolet and optical Telescope. All
the $Swift$-UVOT observations of Mrk 421
available at the HEASARC data archive  were included in our analysis.
The photometry was
computed using a 8$^{\prime\prime}$ source region centered on the Mrk 421 position.
The background was extracted from an annular region (with radius of 20$^{\prime\prime}$ $-$ 50$^{\prime\prime}$)  centered on the source position. The
flux has been corrected for the Galactic extinction using the \cite{card89} parameterization,
with  E$_{B-V}$ = 0.013 mag.

The OVRO
is a 40-m radio telescope working at 15.0 GHz with 3 GHz bandwidth. Mrk 421 was observed by OVRO as part of the blazar monitoring programme  since the end of 2007.
The light curve for Mrk 421 is
publicly available \footnote{http://www.astro.caltech.edu/ovroblazars/} and is directly used in this work.

\section{Results}

\subsection{Light curves}
Figure 1 shows the light curves of Mrk 421  from  radio  to the TeV band.
The time integration is chosen taking into account the sensitivity of the instruments.
For ARGO-YBJ each point corresponds to one month (30 days) of data,
while for $Fermi$-LAT, $Swift$-BAT, $RXTE$-ASM and $MAXI$-GSC the data are averaged over one
week. For $Swift$-XRT and $Swift$-UVOT each point is the result of each dwell, which lasts about hundreds
of seconds.  For OVRO each point is the result of each observation.

\begin{figure}
\includegraphics[width=6in,height=6in]{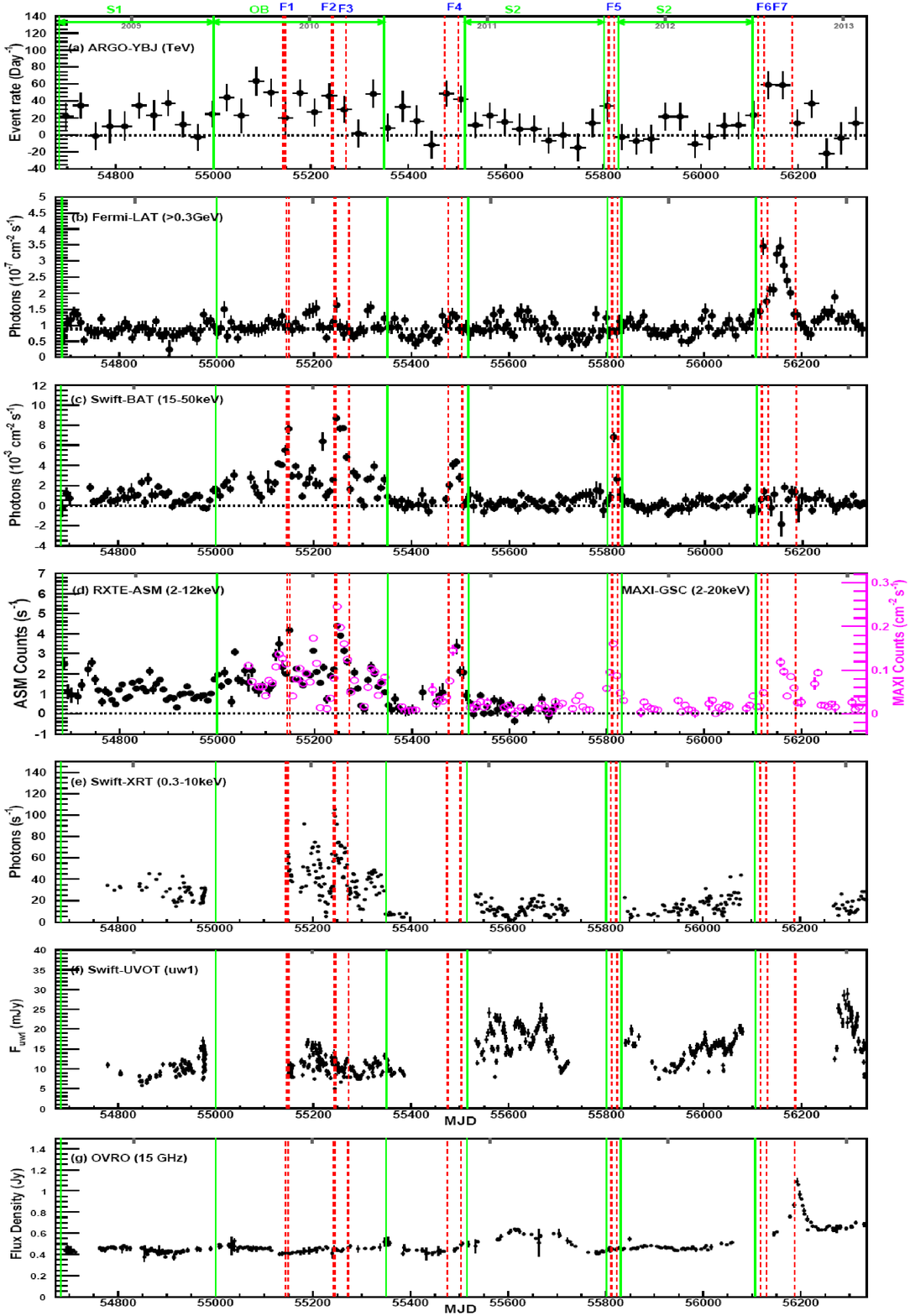}

\caption{
Mrk 421 light curves in different energy bands, from 2008 August 5 to 2013 February 7.
Each bin of the ARGO-YBJ data contains the event rate averaged over 30 days.
Each bin of the  $Fermi$-LAT, $Swift$-BAT, $RXTE$-ASM and $MAXI$-GSC data
contains the event rate averaged over 7 days.  The horizontal dotted line in panel (b) indicates the average flux before F6 and the horizontal dotted lines in the other panels indicate the zero flux.
}
\label{fig1}
\vspace*{-0.5cm}
\end{figure}

In this paper  we will focus on the large X-ray and GeV $\gamma$-ray flares, with the aim  to investigate the spectral variation in different wavebands, compared to the low activity states. We will define different states of activity for Mrk 421 mainly basing on the light curves of $Fermi$-LAT and $Swift$-BAT, partially taking into account the curves of $RXTE$-ASM and $MAXI$-GSC.
For X-ray flares, we will only select the flares which show a large increase both in hard and soft X-rays. According the light curves shown in Figure 1, 5 large X-ray flares (denoted as F1 to F5) and 2 large GeV flares (denoted as F6 and F7) are identified. Beside these flares, we select a long-term outburst phase (denoted as OB)  starting in June 2009  and ending in June 2010, two steady phase (denoted as S1 and S2) with a low activity at all wavebands, i.e. from August 2008  (MJD=54683)  to June 2009  (MJD=55000), and from November 2010 (MJD=55516) to June 2012 (MJD=56106). The duration of all the selected states are indicated in Figure 1.

To estimate the possible observation time for Cherenkov detector during
the same period, a dummy  instrument
located at 30$^{\circ}$  N (close to the latitude of VERITAS, 32$^{\circ}$ N, and MAGIC, 28$^{\circ}$  N) is here considered.
Figure 2 shows the allowable time, requiring
the Sun zenith angle be greater than 105$^{\circ}$, the Moon zenith angle greater than 100$^{\circ}$, and the Mrk 421 zenith angle less than 50$^{\circ}$.
Among the 7 flares presented in this work, F4, F5, F6 and F7 occurred during the period from July to October, forbidden for Cherenkov detector observations, since Mrk 421 is close to the direction of the Sun. The moonlight completely hampered Cherenkov telescope observations during F1 and
partially during F3. Only Flare 2 could have been   observed every night by Cherenkov detectors. Actually VERITAS observed Mrk 421 during the last day of F2 \cite{fort12}. Both VERITAS and HESS observed the first three days of flare F3 \cite{fort12,tlucz11}.

\begin{figure}
\includegraphics[width=6in,height=2.in]{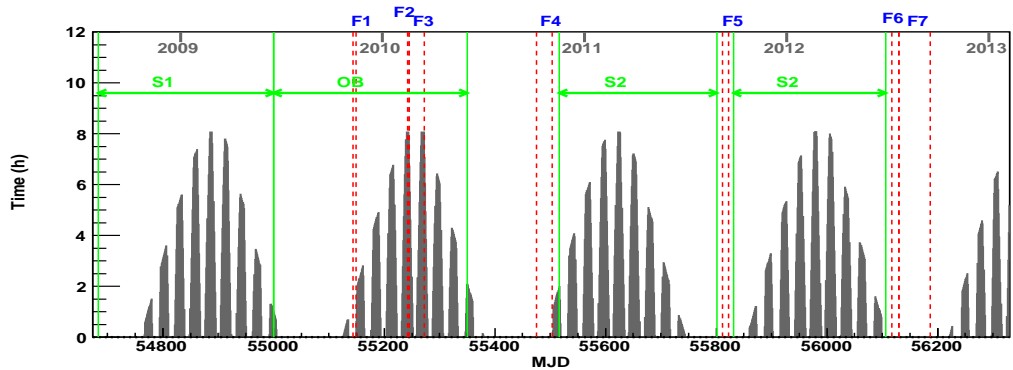}
\caption{
The daily possible Mrk 421 observation time for a Cherenkov detector located at 30$^{\circ}$ North of latitude, during the 4.5 years considered in this work.
}
\label{fig9}
\vspace*{-0.5cm}
\end{figure}

\subsection{Spectral Energy Distribution}
To model the spectral energy distribution, we assume a simple power law spectrum
for $Swift$-BAT, $RXTE$-ASM, $MAXI$-GSC, $Fermi$-LAT and ARGO-YBJ, while for $Swift$-XRT we assume
a logparabolic function.
The time-averaged SEDs for the different activity states, obtained by fitting the data of all the
experiments, are  shown in Figure 3.

According to the $Swift$-XRT data at 0.3$-$10 keV, the peak energy of the
first SED component $E_{peak}$ is  0.394$\pm$0.003 keV and 0.771$\pm$0.003 keV during S2 and S1, respectively.
It increases to 1.429$\pm$0.004 keV during the OB phase and even up to 2.4$-$5.1 keV during F1, F2, and
F3.

According to $MAXI$-GSC measurements at 2$-$20 keV, the flux
during the flaring periods increases by about a factor 4$-$20, compared to S2.
The spectral indexes for most of
the states are in the 2$-$2.4 range while the spectrum softens (index 2.97$\pm$0.18) during F7.
 In the X-ray band (14$-$195 keV)  the $Swift$-BAT data shows an even larger variation (4$-$70
times) with spectral indexes ranging from 2.5 to 3.1.

In the GeV $\gamma$-ray band, the F3, F4, F5, OB, S1 and S2 phases
have similar spectral indexes  and fluxes.
Compared to the S2 state, the spectral index of F1 shows a moderate hardening  (
$\Delta \alpha=$0.20$\pm$0.15) and a flux decrease of  (34$\pm$21)\%.  The spectral index of F2 hardens more significantly ($\Delta \alpha=$0.27$\pm$0.08) with a flux increase of  a factor 2.06$\pm$0.36.
A flux enhancement by a factor 3 is observed during F6 and F7, with a harder
spectral index during F6 ($\Delta \alpha=$0.09$\pm$0.04) and a negligible index variation during F7.

In VHE $\gamma$-ray band the S2 flux is estimated to be (0.33$\pm$0.10) I$_{crab}$.
This result is comparable to the baseline flux of Mrk 421 obtained using a 20 years
long-term combined ACT data \cite{tlucz10}, which is estimated to be less than 0.33 I$_{crab}$
 above 1 TeV.
The averaged measured flux is (0.56$\pm$0.13) and (0.91$\pm$0.14) I$_{crab}$  during S1 and OB phase,
respectively. F2 is the largest flare, achieving a flux of (7.2$\pm$1.5) I$_{crab}$.
The flux of the remaining flares is around (1$-$3) I$_{crab}$.
The spectral index of F7 ($\alpha=$3.22$\pm$0.24) marks the softest spectrum of the observed flares.
The flux modulations appears in coincidence with the X-ray observations.

Summarizing the above results, we can conclude that the flux enhancements are detected in both X-rays
and VHE $\gamma$-rays during all the nine states, compared to the baseline S2. The behavior in the GeV
band is actually different. Accordingly, a phenomenological classification of 3 types of SEDs (T1,
T2 and T3) is here introduced, i.e.,
(I) flares with no or little GeV flux and photon index variations, (II) flares with $\gamma$-ray spectral hardening, irrespective of the flux variations, and (III) flares with flux enhancements, irrespective of spectral behavior.
Type T1 includes phases S1, S2, F3, F4, F5 and OB.
Type T2 includes the F1 and F2 states and also the
day (MJD=56124), corresponding to the F6 maximum flux, during which the spectral index
significantly hardens to $\alpha=$1.60$\pm$0.04.
Type T3 includes the F7 phase. Actually the spectral index becomes softer above the peak energy
for both low and high energy component. The previous flare on May 7th, 2008, reported by
\cite{accia09b}, not included in this present discussion, may also belong to this type.

During flares of types T1 and T2, the peak energies of both the low and high energy components shift to higher energy with respect to the baseline state S2.  This tendency is consistent with most of the previous measurements. This indicates that the modulation of Mrk 421 flux
follows these types in most of the cases.  During flares of type T3, the peak energies could shift to lower energy with respect to S2, but this must be determined by future observations of similar flares.

\begin{figure*}
\includegraphics[width=6in,height=6.5in]{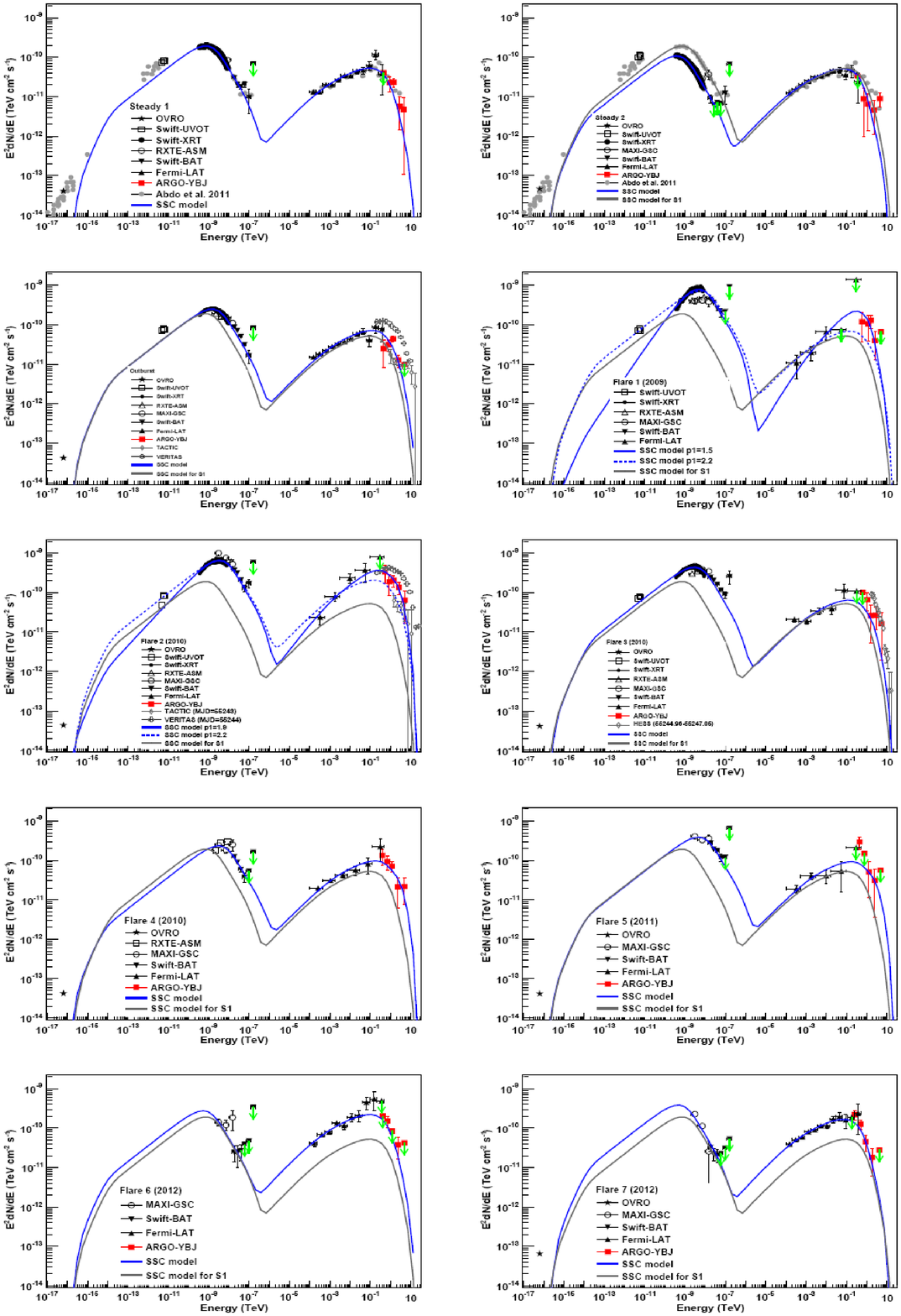}
\caption{
Spectral energy distribution of Mrk 421 during 10 states. The solid line shows the best fit to the data, assuming a homogeneous one-zone SSC model (the best-fit parameters are listed in Table 1).
For comparison, the model describing the   Steady 1 (S1) is also plotted in the other
9 states.
}
\label{fig7}
\vspace*{-0.5cm}
\end{figure*}
\section{The SSC model for Mrk 421}
In this paper the one-zone homogeneous SSC model proposed by \cite{krawczy04} is
adopted to fit the multi-wavelength SEDs measured during different states. In this model a spherical
blob of plasma with a co-moving radius $R$ is assumed. The emission volume is filled with an isotropic population of electrons and a randomly oriented
uniform magnetic field $B$. 
The SED of the injected electrons in the jet frame is assumed to follow a
broken power law with indexes $p1$ and $p2$ below and above the break energy $E_{break}$.
To reduce the free parameters in the model, the low limit for the electron energy $E_{min}$ is fixed to be 500$\cdot m_{e}c^{2}$
 and the high limit $E_{max}$ is assumed to be 10$\cdot E_{break}$.
The radius $R$ for
all phases are arbitrarily set to be 10$^{14}$ m,  being $t_{var}>4.8(20/\delta)$ hours the allowed time
variability range, where $\delta$ is the relativistic Doppler Factor of the
emitting plasma.
According to the SED of flare F1
shown in Figure 3, the ultraviolet and X-ray data cannot be fitted together with a unique
component. Therefore,  the ultraviolet and radio
data points were not used for the fit in this work.
The Extragalactic Background Light (EBL)
absorption on the VHE $\gamma$-ray is included, according to the \cite{franc08} model.

We found that the SSC model  well describes the 10 activity states
considered here, as shown in Fig.3. The found parameters of the model are
given in Table 1.

\begin{table}
\centering
\caption{Best-Fit parameters in the SSC Model  for 10 states}
\label{tab-1}
\begin{tabular}{c|ccc|cc|cc}
\hline
  State& $p_1$  & $p_2$    & log(E$_{break}$) & $\delta$ & B  & $u_{e}$ ($\times 10^{-3}$)               & $u_{e}/u_{B}$ $^{a}$ \\
                &        &          &  (eV)            &          &(G) &(ergs cm$^{-3}$) &  \\
\hline
Steady 1&  2.2  &4.7 & 10.94   &25   &0.093   &7.6    &22.0  \\
Steady 2&  2.2  &4.7 & 10.80   &15   &0.163   &13.2   &12.5   \\
\hline
Outburst&  2.2  &4.7 & 11.11   &25   &0.085   &9.0    &31.6  \\
Flare 1 &  2.2  &4.7 & 11.24   &25   &0.184   &4.5    &3.3  \\
Flare 2 &  2.2  &4.7 & 11.27   &25   &0.100   &12.7   &31.7  \\
Flare 3 &  2.2  &4.7 & 11.16   &25   &0.136   &5.7    &7.8  \\
Flare 4 &  2.2  &4.7 & 11.38   &25   &0.052   &14.2   &135  \\
Flare 5 &  2.2  &4.7 & 11.37   &25   &0.082   &9.6    &36.2  \\
Flare 6 &  2.2  &4.7 & 10.94   &25   &0.068   &20.3   &110  \\
Flare 7 &  2.2  &4.7 & 10.79   &25   &0.119   &11.9   &21.2  \\
\hline
Flare 1 &  1.5  &4.7 & 11.31   &10   &0.202   &25.7   &15.9  \\
Flare 2 &  1.9  &4.7 & 11.33   &20   &0.073   &21.0   &100  \\
\hline
\multicolumn{8}{l}{$^{a}$ $u_{e}/u_{B}$=8$\pi$$u_{e}$/B$^{2}$}
\end{tabular}
\end{table}

For most of the states, the SED can be reasonably described using
injected electrons with  $p1$=2.2, i.e. the canonical particle spectral index
predicted for relativistic diffuse shock acceleration. This
suggests that this process is active in Mrk 421. According to SSC modeling result, the variation of type T1 and T3 are mainly caused by variation of the break energy $E_{break}$, which may be due to the variation of the ambient medium  that is crossed by the shock.
The T2 flares require a harder injected electron spectrum than the other types. This change would be caused by the acceleration processes. Some possible choices include the relativistic shock with extreme parameters\cite{steck07}, the particles re-accelerated by the stochastic process in the downstream region\cite{virtan05}, particle accelerated within a relativistic magnetic reconnection\cite{guo14}.

\section{Summary }
In conclusion, we have presented a 4.5-year long-term continuous simultaneous multi-wavelength monitoring of Mrk 421 in the whole
emission range from radio to TeV gamma rays. The observation time covers both active and steady states. For the first time, the SED evolutions of Mrk 421 during different states, compared to the baseline state,  have been achieved.  The SEDs of different states can be reasonably described by a simple one-zone SSC model. The flux and spectral variations in the different states may be caused by
different intrinsic mechanism related to the acceleration process or to variations of the ambient medium.

\end{document}